\documentclass[conference, 10pt]{IEEEtran}

\usepackage[utf8]{inputenc}
\usepackage[T1]{fontenc}

\usepackage{graphicx}
\usepackage{color}
\usepackage{placeins}
\usepackage{float}
\usepackage{hyperref}
\usepackage{url}
\usepackage[cmex10]{amsmath}
\usepackage{amssymb,amsthm}
\usepackage{bm}

\usepackage{tikz}
\usetikzlibrary{calc}

\usepackage[backend=biber,style=ieee,mincrossrefs=1000]{biblatex}
\AtBeginDocument{\toggletrue{blx@useprefix}}
\AtBeginBibliography{\togglefalse{blx@useprefix}}

\DeclareMathOperator{\rect}{rect}
\DeclareMathOperator{\round}{round}

\newcommand{\E}{\operatorname{\mathbb E}}

\newcommand{\Z}{\operatorname{\mathbb Z}}

\allowdisplaybreaks[2]
\graphicspath{{figures/}}

\begin{document}
%
\title{End-to-end optimization of nonlinear\\transform codes for perceptual quality}

\author{\IEEEauthorblockN{Johannes Ballé, Valero Laparra, Eero P. Simoncelli}
\IEEEauthorblockA{Center for Neural Science and
Courant Institute of Mathematical Sciences\\
New York University,
New York, NY, USA\\
\{johannes.balle,valero,eero.simoncelli\}@nyu.edu}}



\maketitle

\begin{abstract}
We introduce a general framework for end-to-end optimization of the rate--distortion performance of nonlinear transform codes assuming scalar quantization. The framework can be used to optimize any differentiable pair of analysis and synthesis transforms in combination with any differentiable perceptual metric. As an example, we consider a code built from a linear transform followed by a form of multi-dimensional local gain control. Distortion is measured with a state-of-the-art perceptual metric. When optimized over a large database of images, this representation offers substantial improvements in bitrate and perceptual appearance over fixed (DCT) codes, and over linear transform codes optimized for mean squared error.
\end{abstract}



%
\IEEEpeerreviewmaketitle

\section{Introduction}
\begin{tikzpicture}[remember picture, overlay]
\node[below left,xshift=-1ex,yshift=-1ex] at (current page.north east) {
Accepted as a conference contribution to Picture Coding Symposium 2016 \copyright{} IEEE
};
\end{tikzpicture}
Transform coding~\cite{Go01} is one of the most successful areas of signal processing. Virtually all modern image and video compression standards operate by applying an invertible transformation to the signal, quantizing the transformed data to achieve a compact representation, and inverting the transform to recover an approximation of the original signal.

Generally, these transforms have been linear. Non-Gaussian/nonlinear aspects of signal statistics are typically handled by augmenting the linear system with carefully selected nonlinearities (for example, companding nonlinearities to enable non-uniform quantization, prediction for hybrid compression, etc.). Deciding which combination of these operations, also known as “coding tools,” are ultimately useful is a cumbersome process. The operations are generally studied and optimized individually, with different objectives, and any proposed combination of coding tools must then be empirically validated in terms of average code rate and distortion.

This is reminiscent of the state of affairs in the field of object and pattern recognition about a decade ago.  As in the compression community, most solutions were built by manually combining a sequence of individually designed and optimized processing stages. In recent years, that field has seen remarkable performance gains~\cite{DeDoSoLiLi09}, which have arisen primarily because of end-to-end system optimization.  Specifically, researchers have chosen architectures that consist of a cascade of transformations that are differentiable with respect to their parameters, and then used modern optimization tools to jointly optimize the full system over large databases of images.

\begin{figure}
\centering\noindent%
\includegraphics[width=\columnwidth]{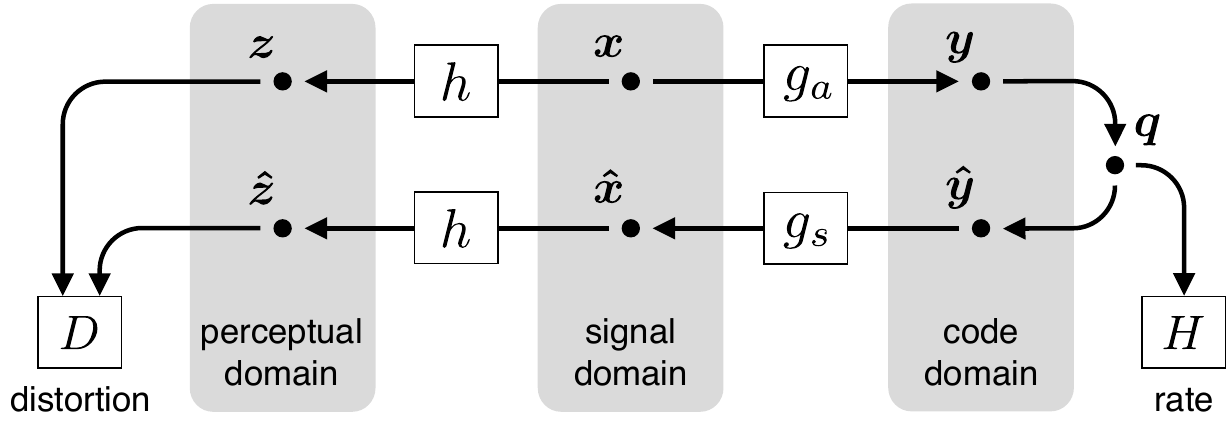}%
\vspace*{-1ex}
\caption{Nonlinear transform coding optimization framework.  See text.}\label{fig:overview}
\end{figure}

Here, we take a step toward using such end-to-end optimization in the context of compression. We develop an optimization framework for nonlinear transform coding (fig.~\ref{fig:overview}), which generalizes the traditional transform coding paradigm. An image vector $\bm x$ is transformed to a code domain vector using a differentiable function $\bm y=g_a(\bm x; \bm \phi)$ (the analysis transform), parameterized by a vector $\bm \phi$ (containing linear filter coefficients, for example). The transformed $\bm y$ is subjected to scalar quantization, yielding a vector of integer indices $\bm q$ and a reconstructed vector $\bm{\hat y}$. The latter is then nonlinearly transformed back to the signal domain to obtain the reconstructed image $\bm{\hat x} = g_s(\bm{\hat y}; \bm \theta)$, where this synthesis transform $g_s$ is parameterized by vector $\bm \theta$.

The code rate is assessed by measuring the entropy, $H$, of the discrete probability distribution $P_{\bm q}$ of the quantization indices over an ensemble of images. Traditionally, the distortion is assessed directly in the image domain by taking the squared Euclidean norm of the difference between $\bm x$ and $\bm{\hat x}$ (or equivalently, the peak signal-to-noise ratio, PSNR). However, it is well known that PSNR is not well-aligned with human perception~\cite{Gi93}. To alleviate this problem, we allow an additional “perceptual” transform of both vectors $\bm z=h(\bm x)$ and $\bm{\hat z}=h(\bm{\hat x})$, on which we then compute distortion using a suitable norm. A well-chosen transform~$h$ can provide a significantly better approximation of subjective visual distortion than PSNR (e.g.,~\cite{LaBaBeSi16}).

\section{Optimization framework}
\label{sec:framework}
In the transform coding framework given above, we seek to adjust the analysis and synthesis transforms $g_a$ and $g_s$ so as to minimize the rate--distortion functional:
\begin{equation}
\label{eq:L}
L[g_a,g_s] = H[P_{\bm q}] + \lambda \E \|\bm z - \bm{\hat z}\|.
\end{equation}
The first term denotes the discrete entropy of the vector of quantization indices $\bm q$. The second term measures the distortion between the reference image $\bm z$ and its reconstruction $\bm{\hat z}$ in a perceptual representation. Note that both terms are expectations taken over an ensemble of images.

We wish to minimize this objective over the continuous parameters $\{\bm\theta, \bm\phi\}$. Most optimization methods rely on differentiability, but both terms in the objective depend on the quantized values in $\bm q$, and the derivative of the quantizer is discontinuous (specifically, it is zero or infinite everywhere). To resolve this, we propose to approximate the objective function with one that is continuously differentiable, by replacing the deterministic quantizer with an additive uniform noise source.

\begin{figure}
\centering\noindent%
\includegraphics[width=\linewidth,trim=0 12 0 0]{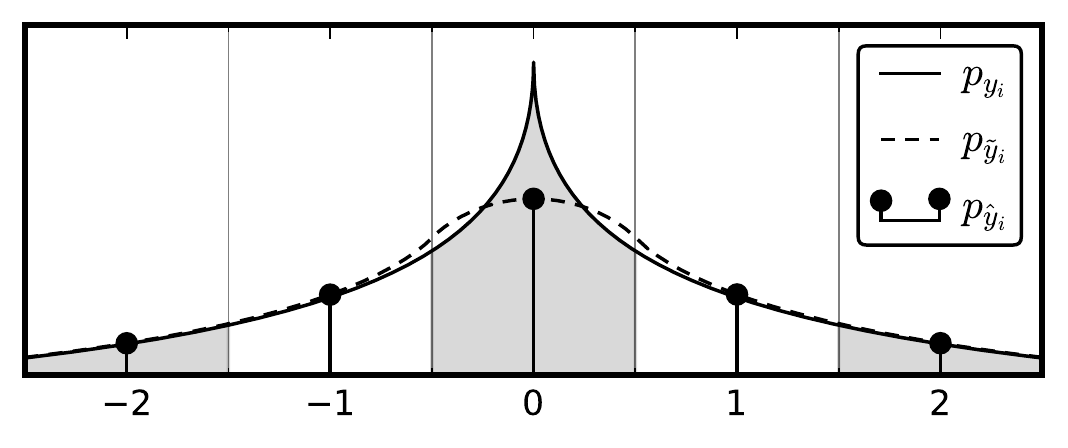}%
\caption{Relationship between densities of $y_i$, $\tilde y_i$, and $\hat y_i$. $p_{\tilde y_i}$ is a continuous relaxation of the the probability masses in each of the quantization bins.}\label{fig:dithering}
\end{figure}

A uniform scalar quantizer is a piecewise constant function applied to each of the elements of $\bm y$: $\hat y_i = \round(y_i)$.\footnote{Without loss of generality, we assume that the quantization bin size is~1, since we can always modify the analysis/synthesis transforms to include a rescaling. Further, we can implement non-uniform quantization by using nonlinear transforms (as in \emph{companding}).} The marginal density of the quantized values is given by:
\begin{equation}
p_{\hat y_i}(t) = \sum_{n=-\infty}^{\infty} P_{q_i}(n)\, \delta(t-n),
\label{eq:sampling}
\end{equation}
where
\begin{equation}
P_{q_i}(n) = (p_{y_i} \ast \rect)(n), \text{ for all } n \in \Z,
\end{equation}
is the probability mass in the $n$th quantization bin. Here, ‘$\ast$’ represents continuous convolution, and $\rect$ is a uniform distribution on $(-\frac 1 2, \frac 1 2)$. If we add independent uniform noise to $y_i$, i.e., form the signal $\tilde y_i = y_i + \Delta y_i$, with $\Delta y_i \sim \rect$, then the density of that signal is $p_{\tilde y_i} = p_{y_i} \ast \rect$. $p_{\tilde y_i}$ is identical to $P_{q_i}$ at all integer locations, and provides a \emph{continuous relaxation} for intermediate values (fig.~\ref{fig:dithering}). We propose to optimize the differential entropy $h[ p_{\tilde y_i} ]$ as a proxy for the discrete entropy $H[P_{q_i}]$. To optimize it, we need a running estimate of $p_{\tilde y_i}$. This estimate need not be arbitrarily precise, since $p_{\tilde y_i}$ is band-limited by convolution with $\rect$. Here, we simply use a non-parametric, piecewise linear function (a first-order spline approximation). We also use $\tilde y_i$ rather than $\hat y_i$ to obtain gradients of the distortion term. The overall objective can be written as:
\begin{multline}
L(\bm \theta, \bm \phi) = \E_{\bm x,\Delta \bm y}\Bigl( - \log_2 p_{\bm{\tilde y}}(g_a(\bm x; \bm \phi) + \Delta \bm y) \\
+ \lambda\, \bigl\|h\bigl(g_s(g_a(\bm x; \bm \phi) + \Delta \bm y; \bm \theta)\bigr) - h(\bm x)\bigr\| \Bigr),
\end{multline}
where $p_{\bm{\tilde y}}(\bm{\tilde y}) = \prod_i p_{\tilde y_i}(\tilde y_i)$. This is differentiable with respect to $\bm \theta$ and $\bm \phi$ and thus suited for stochastic gradient descent. Although finding a global optimum is not guaranteed, this optimization problem is similar to others which are encountered when optimizing deep neural networks, and which have been found to behave well in practice.

\section{Choice of parametric transforms}
\label{sec:transforms}
In a traditional transform code, both analysis and synthesis transforms are linear, and exact inverses of each other. In general, this need not be the case, so long as the overall system minimizes the rate--distortion functional. We have previously shown that a linear transform followed by a particular form of joint local gain control (generalized divisive normalization, GDN) is well-matched to the local probability structure of photographic images~\cite{BaLaSi15}. This suggests that jointly normalized representations might also prove useful for compression. To demonstrate the use of our optimization framework, we examine GDN as a candidate analysis transform, and introduce an approximate inverse as the corresponding synthesis transform. For the perceptual transform, we use the normalized Laplacian pyramid~\cite{LaBaBeSi16} (NLP), which mimics the local luminance and contrast behaviors of the human visual system.

\subsection{Generalized divisive normalization (GDN)}
The GDN transform consists of a linear decomposition $\bm H$ followed by a joint nonlinearity, which divides each linear filter output by a measure of overall filter activity:
\begin{align}
\label{eq:gdn}
\bm y = g_a(\bm x; \bm \phi)
  && \text{s.t.} && y_i &= \frac {v_i} {\bigl(\beta_i + \sum_j \gamma_{ij}|v_j|^{\alpha_{ij}}\bigr)^{\varepsilon_i}} \notag \\
  && \text{and} && \bm v &= \bm H \bm x,
\end{align}
with parameter vector $\bm \phi = \{\bm \alpha, \bm \beta, \bm \gamma, \bm \varepsilon, \bm H\}$.

\subsection{Approximate inverse of GDN}
The approximate inverse we introduce here is based on the fixed point iteration for inversion of GDN introduced in \cite{BaLaSi15}. It is similar in spirit to the LISTA algorithm~\cite{GrLe10}, in that it uses the parametric form of the inversion iteration, but unties its parameters from their original values for faster convergence. We find that for purposes of image compression, one iteration is sufficient:
\begin{align}
\label{eq:invgdn}
\bm{\hat x} = g_s(\bm{\hat y}; \bm \theta)
  && \text{s.t.} && \bm{\hat x} &= \bm H' \bm w \notag \\
  && \text{and} && w_i &= \hat y_i \cdot \bigl(\beta_i' + \sum_j \gamma_{ij}'|\hat y_j|^{\alpha_{ij}'}\bigr)^{\varepsilon_i'},
\end{align}
where the parameter vector consists of a distinct set of parameters: $\bm \theta = \{\bm \alpha', \bm \beta', \bm \gamma', \bm \varepsilon', \bm H'\}$.

\subsection{Normalized Laplacian pyramid (NLP)}
The NLP imitates the transformations associated with the early visual system: local luminance subtraction and local gain control~\cite{LaBaBeSi16}. Images are decomposed using a Laplacian pyramid~\cite{BuAd83}, which subtracts a local estimate of the mean luminance at multiple scales. Each pyramid coefficient is then divided by a local estimate of amplitude (a constant plus the weighted sum of absolute values of neighbors). Perceptual quality is assessed by evaluating a norm of the difference between reference and reconstruction in this normalized domain. The parameters (constant and weights used for amplitudes) are optimized to best fit perceptual data in the TID2008 database~\cite{PoLuZeEgCa09}, which includes images corrupted by artifacts arising from compression with block transforms. This simple distortion measure provides a near-linear fit to the human perceptual judgments in the database, outperforming the widely-used SSIM~\cite{WaSiBo03} and MS-SSIM~\cite{WaBoShSi04} quality metrics~\cite{LaBaBeSi16}. Examples and code are available at \url{http://www.cns.nyu.edu/~lcv/NLPyr}.

\section{Experimental results}
\label{sec:experiments}

\begin{figure}
\centering\noindent%
\includegraphics[width=\linewidth,trim=0 10 0 0]{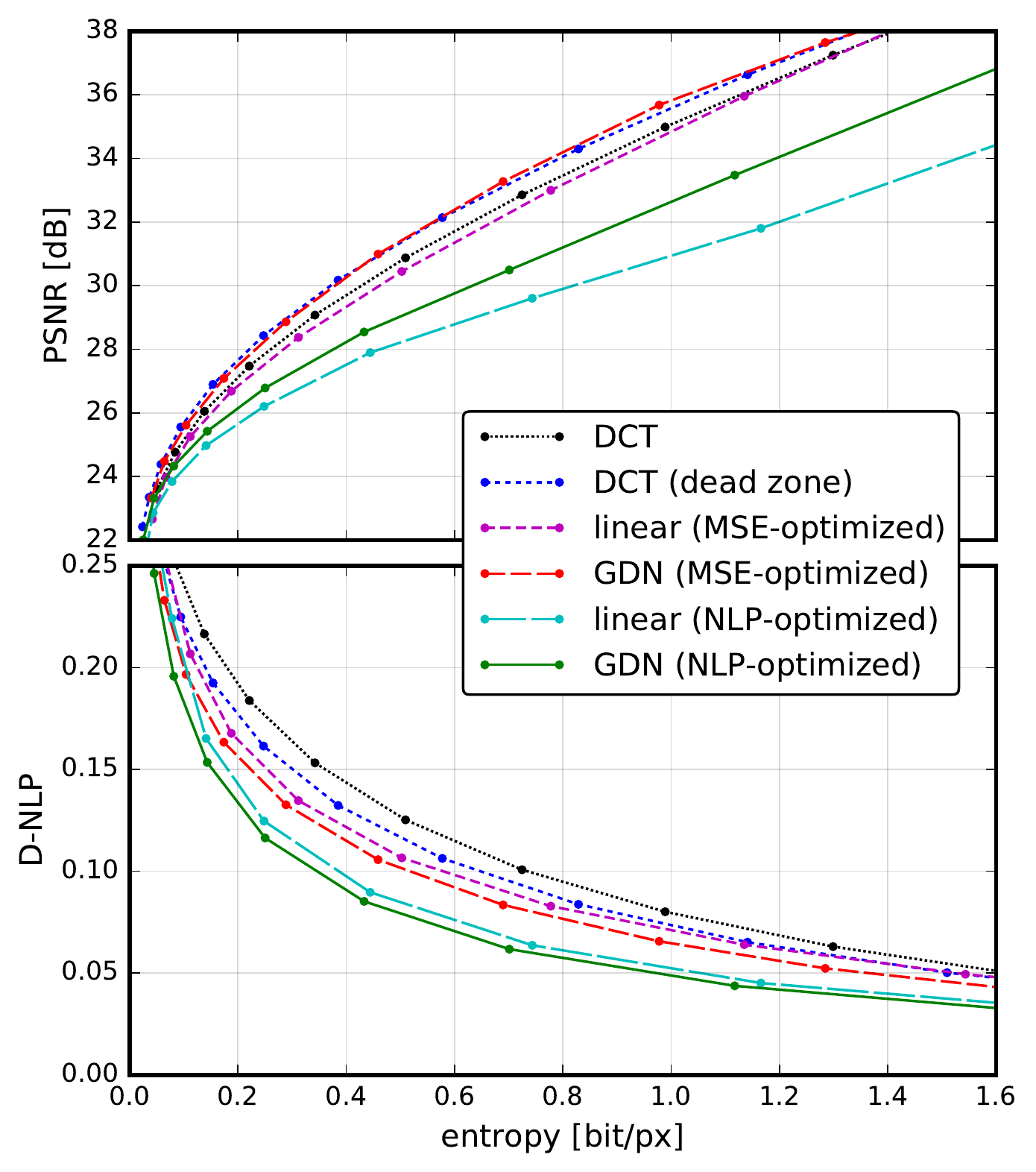}%
\caption{Rate--distortion results averaged over the entire Kodak image set (24 images, $752 \times 496$ pixels each). Reported rates are average discrete entropy estimates. Reported distortion is average PSNR (top) and distance in the normalized Laplacian pyramid domain (bottom -- see text).}\label{fig:rd}
\end{figure}

The proposed framework can be used to optimize any differentiable pair of analysis and synthesis transforms in combination with any differentiable perceptual metric. Here, we consider two types of transform: a linear analysis and synthesis transform operating on $16\times 16$ pixel blocks (in this case, $\bm \theta$ and $\bm \phi$ each only consist of $256 \times 256$ filter coefficients), and a $16\times 16$ block GDN transform with the approximate inverse defined above (with $\bm \theta$ and $\bm \phi$ consisting of filter coefficients and normalization parameters, as defined above). We optimized each of these for two distortion metrics: mean squared error (MSE) and distance in the NLP domain~\cite{LaBaBeSi16}.  Each combination of transform and distortion metric was optimized for different values of $\lambda$. We also include a fixed linear transform, the $16\times 16$ discrete cosine transform (DCT), with or without dead-zone quantization, serving as a baseline. All other codes (i.e., those optimized in our framework) are constrained to use uniform quantization.

We used the Adam algorithm~\cite{KiBa14}, a variant of stochastic gradient descent, to optimize all codes over a large collection of images from the ImageNet database~\cite{DeDoSoLiLi09}, initializing the parameters randomly. For each optimization step, we used a randomly selected mini-batch of 4 images of $128\times 128$ pixels. To prevent overfitting to the training database, we performed all evaluations on a separate set of test images.\footnote{The Kodak image set, downloaded from \url{http://www.cipr.rpi.edu/resource/stills/kodak.html}. We converted the images to grayscale and discarded 8 pixels from each side to eliminate boundary artifacts.}

We measured rate--distortion performance for all four transform/distortion metric combinations, along with the DCT transform. Note that the additive noise approximation was used only for optimization, not for evaluation: We evaluated rates by estimating discrete entropy of the quantized code vector $\bm q$. For each choice of $\lambda$, we optimized a separate set of transform parameters, which could be stored in the encoder and decoder. The only side information a real-world codec would need to transmit is the choice of $\lambda$ and the image size (although it would be desirable to reduce the storage requirements by storing parameters jointly for different $\lambda$).

\begin{figure}
\centering\noindent\footnotesize%
\includegraphics[width=\columnwidth,trim=0 45 0 0,clip]{{{images/00009/16x16-blocks-nlp-3-gdn/00003}}}\\%
{\bf NLP-GDN}, 0.190 bit/px. PSNR: 20.95 D-NLP: 0.21 MS-SSIM: 0.868\\%
\vspace{2mm}%
\includegraphics[width=\columnwidth,trim=0 45 0 0,clip]{{{images/00009/16x16-blocks-fixed-dct/00013}}}\\%
{\bf DCT (dead z.)}, 0.204 bit/px. PSNR: 21.81 D-NLP: 0.28 MS-SSIM: 0.827\\%
\vspace{2mm}%
\includegraphics[width=\columnwidth,trim=0 45 0 0,clip]{{{images/00009/original}}}\\%
{\bf Original}, 8 bit/px. PSNR: $\infty$ D-NLP: 0 MS-SSIM: 1%
\caption{Example image from Kodak set (bottom), compressed with DCT and hand-optimized (for MSE) dead-zone quantization (middle), and GDN with uniform quantization optimized in the NLP domain (top). Cropped to fit page.}\label{fig:visual1}
\end{figure}

\begin{figure}
\centering\noindent\footnotesize%
\includegraphics[width=\columnwidth,trim=0 25 0 20,clip]{{{images/00011/16x16-blocks-nlp-3-gdn/00001}}}\\%
{\bf NLP-GDN}, 0.044 bit/px. PSNR: 26.37 D-NLP: 0.21 MS-SSIM: 0.881\\%
\vspace{2mm}%
\includegraphics[width=\columnwidth,trim=0 25 0 20,clip]{{{images/00011/16x16-blocks-fixed-dct/00007}}}\\%
{\bf DCT (dead z.)}, 0.044 bit/px. PSNR: 26.93 D-NLP: 0.24 MS-SSIM: 0.857\\%
\vspace{2mm}%
\includegraphics[width=\columnwidth,trim=0 25 0 20,clip]{{{images/00011/original}}}\\%
{\bf Original}, 8 bit/px. PSNR: $\infty$ D-NLP: 0 MS-SSIM: 1%
\caption{A second example image from the Kodak set (see caption for fig. \ref{fig:visual1}).}\label{fig:visual2}
\end{figure}

For evaluation of the distortion, we first computed the mean squared error (MSE) over the entire image set for each $\lambda$, and then converted these values into PSNRs (fig.~\ref{fig:rd}, top panel). In terms of PSNR, the optimized linear transform is slightly worse than the DCT, because the statistics of the ImageNet database are slightly different from the Kodak set.\footnote{If we validate on a held-out test set of images from ImageNet, the two transforms perform equally well.} The DCT with dead-zone quantization is better, but doesn't outperform the MSE-optimized GDN transform, which uses only uniform quantization. The NLP-optimized transforms don't perform well in terms of PSNR.

The situation is reversed, however, when we examine performance in terms of perceptual distortion (fig.~\ref{fig:rd}, bottom). Here, we evaluated the norm in the NLP domain (D-NLP) for each image in the set, and then averaged across images. Note that this norm is almost (inversely) proportional to  human perceptual mean opinion scores (MOS) across several image databases~\cite{LaBaBeSi16}. Overall, the combination of NLP and GDN achieves an impressive rate savings at similar quality when compared with MSE-optimized methods, and with the DCT (both uniform and dead-zone quantizers).  It is also interesting to note that in terms of NLP distance, the optimized linear transform with uniform quantization outperforms both versions of the DCT. This may be because the optimized filters tend to be spatially localized (and oriented/bandpass), which possibly leads to visually less disturbing artifacts (not shown).

For visual evaluation, we  show results on two example images (figs. \ref{fig:visual1}~and~\ref{fig:visual2}). Results for the entire test set are available at \url{http://www.cns.nyu.edu/~balle/nlpgdn}. The figures serve to illustrate the main effect of using a perceptual metric that is aware of local \emph{relative} contrast. Traditional, linear systems optimized for MSE give too much preference to high-contrast regions (e.g., the snow-covered mountains in the background, or the pebbles/debris in the foreground; fig.~\ref{fig:visual1}, center image). By performing joint normalization before quantization, the NLP-optimized GDN transform allocates more bits to represent detail in low-contrast regions (such as the forest in the depicted scene; top image). Overall, the rate allocation is perceptually more balanced, which leads to a more appealing visual appearance.

\section{Discussion}
\label{sec:discussion}
We have introduced a framework for end-to-end optimization of nonlinear transform codes, which can be applied to any set of parametric, differentiable analysis and synthesis transforms. By optimizing a nonlinear transform for a perceptual metric over a database of photographs, we obtained a nonlinear code that respects the perception of local luminance and contrast errors, allowing for significant rate savings.

The earliest instance of a linear transform optimized for signal properties may be the Karhunen--Loève transform (KLT), or principal components analysis (PCA). The DCT was originally introduced as an efficient approximation to the KLT for a separable autoregressive process of order 1~\cite{AhNaRa74}. Other studies have optimized transform parameters specifically for perceptual compression (e.g.,~\cite{AhPe92,Wa93}), but these were generally limited to optimizing weighting matrices for the DCT. Models that use “matched” nonlinear transformations as a means of converting to/from a more desirable representation of the data are known in the machine learning literature as \emph{autoencoders}~\cite{HiSa06}. However, we are unaware of any work that directly aims to optimize discrete entropy.

We assume uniform quantization in the transform domain, and replace quantization with additive uniform noise to relax the discontinuous problem into a differentiable one.
We find that this method empirically outperforms results obtained by ignoring the effects of the discontinuous quantizer on the backpropagation of gradients (not shown). Under the presented conditions, adding noise is equivalent to performing dithered quantization~\cite{Sc64}. Dithering was used in some early image coders~\cite{Ro62}, but generally does not improve rate--distortion performance. To our knowledge, it has not been used as a form of continuous relaxation for optimization purposes. While uniform quantization has been shown to be asymptotically optimal~\cite{GiPi68}, it is well known that dead-zone quantization generally performs better for linear transform coding of images. Here, we demonstrate empirically that the use of nonlinear transforms with uniform quantization allows equivalent or better solutions, and our framework provides a means of finding these transforms.

Divisive normalization has previously been used in DCT-based image compression, e.g.,~\cite{MaEpNaSi06,WaReWaMaGa13}. These approaches use the normalized representation both for coding and distortion estimation, reasoning that this domain is both perceptually and statistically uniform, and thus well-suited for both. The framework introduced here offers more flexibility, by allowing the perceptual domain and the code domain to be distinct (fig.~\ref{fig:overview}). Further, previous methods required the decoder to invert the normalization transform, either by solving an iterative set of linear  equations for every block~\cite{MaEpNaSi06}, estimating the multipliers (i.e., the values of the denominators) from neighboring blocks~\cite{WaReWaMaGa13}, or embedding the multipliers into the code as side information. Our framework eliminates this problem by introducing a highly efficient approximate inverse transform, which is jointly optimized along with the normalization transform.

There are several directions in which to proceed with this work. Well-known techniques to improve performance of linear transform codes, such as run-length encoding, adaptive entropy coding, and signal-adaptive techniques in general, should be investigated in the context of nonlinear transform coding. Furthermore, our framework offers a means for exploring much more sophisticated nonlinear analysis/synthesis transforms as well as perceptual metrics, since it is built on the highly successful paradigm of end-to-end optimization over training data.

\section*{Acknowledgement}
JB and EPS are supported by the Howard Hughes Medical Institute. VL is supported by the APOSTD/2014/095 Generalitat Valenciana grant (Spain).


\printbibliography

\end{document}